# Anisotropic conjugated polymer chain conformation tailors the energy migration in nanofibers

Andrea Camposeo[†,*], Ryan D. Pensack[#], Maria Moffa[†], Vito Fasano[§], Davide Altamura[‡], Cinzia Giannini[‡], Dario Pisignano[†,§,**], Gregory D. Scholes[#,***]

[†] Istituto Nanoscienze-CNR, Euromediterranean Center for Nanomaterial Modelling and Technology (ECMT), via Arnesano, I-73100, Lecce, Italy.

[#] Department of Chemistry, Princeton University, Princeton NJ 08544 U.S.A.

[‡] Istituto di Cristallografia (IC-CNR), via Amendola 122/O, I-70126, Bari, Italy.

[§] Dipartimento di Matematica e Fisica "Ennio De Giorgi", Università del Salento, via Arnesano, I-73100, Lecce, Italy.

ABSTRACT

Conjugated polymers are complex multi-chromophore systems, with emission properties strongly dependent on the electronic energy transfer through active sub-units. Although the packing of the conjugated chains in the solid state is known to be a key factor to tailor the electronic energy transfer and the resulting optical properties, most of the current solution-based processing methods do not allow for effectively controlling the molecular order, thus making the full unveiling of energy transfer mechanisms very complex. Here we report on conjugated polymer fibers with tailored internal molecular order, leading to a significant enhancement of the emission quantum yield. Steady state and femtosecond time-resolved polarized spectroscopies evidence that excitation is directed toward those chromophores oriented along the fiber axis, on a typical timescale of picoseconds. These aligned and more extended chromophores, resulting from the high stretching rate and electric field applied during the fiber spinning process, lead to improved emission properties. Conjugated polymer fibers are relevant to develop optoelectronic plastic devices with enhanced and anisotropic properties.

## 1. INTRODUCTION

In the framework of enhanced and flexible materials for optoelectronics and energy applications,[1,2] organic semiconductors[3] have been largely exploited due to their favorable charge-carrier and emission properties, and easy processing by solution methods, printing, and soft lithographies.[4-7] Current applications include, but are not limited to, light-emitting devices,[8] photovoltaic cells,[9] field effect transistors[10] and lasers.[11] The optoelectronic properties of organic semiconductors are mainly determined by the behavior of excitons, quasi-particles formed by an electron tightly bounded to a hole, which can be excited either optically or electrically. Such excitations are typically localized on the scale of a single or few molecules, and their migration and recombination dynamics determines both light-emission and light-harvesting properties.[12,13] More specifically, short migration lengths are preferred for high fluorescence yield because this decreases the possibility of quenching excitons by encountering trap states and "concentration quenching".[14-16] Moreover, in photovoltaics long exciton migration lengths might favor exciton dissociation at the interface between organic semiconductors and electron/hole acceptors.[17] Therefore, understanding and tailoring how excitation energy travels in organic semiconductors is an issue for improving device performance.

In fact, most current applications rely on materials with randomly oriented conjugated chains, which might limit the yield of photogenerated charges or photoluminescence.[18] In such systems, electronic energy transfer[19] directs the excitation toward lower energy chromophores, typically associated to low-emissive aggregates. Although several studies [20-22] on isolated conjugated polymer chains suggested that a route to improve charge mobilities and emission yield is the development of effective processing methods which can force macromolecules to adopt extended and elongated conformations, little has been done in this direction. Indeed, individual conjugated polymer chains are intrinsically anisotropic and any high-performing solid state structure must inherit such a property.[20,23,24] For instance, conjugated polymers were aligned in host-guest systems by using mesoporous silica templates,[25] which led to fast intrachain electronic energy transfer,[20] highly polarized emission[26] and low-threshold amplified spontaneous emission.[27] However, these results were limited to templates with a pore size smaller than 5 nm, hosting isolated polymer chains, whereas properties similar to those of spin-cast films are found as soon as the channel diameter increases above 5 nm.[28] In other attempts, an alignment of specific, molecularly designed conjugated polymer chains was induced along an applied flow field,[24] a method allowing for obtaining a three-order of magnitude faster hole mobility along the direction of chain alignment compared to the perpendicular one, as well as high absorption and emission dichroic ratios. Poly[2-methoxy-5-(2-ethylhexyloxy)-1,4-phenylene-vinylene] (MEH-PPV) has also been aligned using liquid crystalline hosts.[29-32] While it was established that the conjugated chains were highly extended, the procedure is only possible for very dilute polymer-liquid crystal solutions. In another approach, a simple shear flow was exploited to induce an extension of the MEH-PPV molecules in solution, using a Coutte cell.[33,34] A small increase of the polarization anisotropy was observed, together with a variation of the emission intensity and wavelength depending on the solution viscosity and solvents, which was attributed to different energy migration pathways activated by shear-induced modification of the molecular conformation. Despite of such efforts, solid state structures made of conjugated polymers are still far from the performances expected for extended chains.

Recently, micro- and nanofibers made by conjugated polymers have emerged as a novel class of nanostructured, solid state materials featuring enhanced properties compared to bulk or thin film samples.[35] Various approaches for producing polymer fibers are currently based on extensional flows or on other methods which induce extended and ordered chain configurations. As a consequence, improved charge mobilities,[23,36] emission quantum yield,[37,38] and polarized emission[38] are observed in these systems. These nanostructures have already demonstrated potential as the active component of organic light-emitting diodes, solar cells, and transistors,[39] and can constitute a valuable benchmark system to tailor and control the fundamental properties of conjugated molecules in the solid state.

Here we report on fibers made of MEH-PPV, realized by electrospinning. Tailoring the degree of order of molecular chains through solution control is revealed by polarization spectroscopies and X-ray diffraction measurements. A comparison with thin films indicates that fibers composed of aligned and more extended polymer chains achieve a fivefold enhancement of the emission quantum yield as well as emission polarized along the fiber axis, independent of excitation configuration. Femtosecond pump–probe anisotropy measurements indicate that, in the ordered fibers, photogenerated excitons preferentially re-orient along the fiber axis. We find that the timescale of the energy migration is of the order of that expected for enhanced interchain coupling.

## 2. RESULTS AND DISCUSSION

**MEH-PPV nanofibers**. The MEH-PPV nanofiber samples used in this work are shown in Figure 1. MEH-PPV is chosen as a prototype conjugated polymer, whose optical properties have been investigated in different packing configurations of the macromolecules, from single molecules,[40] to isolated and stretched individual chains,[21,22] small aggregates[41] and thin films.[42] Uniaxially-oriented arrays of fibers (Fig. S1) are produced by using a rotating collector during electrospinning,[43] then they are collected onto quartz substrates and encapsulated in a photocurable polymer (Fig. 1a, b) to prevent photo-oxidation (Fig. S2). The encapsulation of the fibers in the resin also decreases the amount of reflected light, because of the reduced refractive index contrast with the conjugated polymer. The fibers have a ribbon shape, with an average ratio of their width and thickness of 2.5 (Fig. 1c,d), and an average width of 300 nm. The rapid evaporation of the more volatile component (chloroform) of the liquid jet during the spinning process and the formation of a collapsing solid skin can be the origin of the ribbon shape.[44] A flattening of the fibers upon substrate deposition, due to the presence of a fraction of slowly evaporating solvent (dimethyl sulfoxide, DMSO) in the solution, may also occur.[45]

The possibility of obtaining fibers with strongly anisotropic packing of the constituent polymer chains[46,47] is related to the very high strain rate ($\sim$10$^3$ s$^{-1}$)[48] exerted in electrospinning, which typically induces molecular stretching and orientation in the jet, that is then retained in the fibers. In general, conjugated polymers are poorly viscoelastic,[49] a property often associated with their semi-flexible backbones, and the final degree of chain orientation depends on many factors.[46-48] Modeling of the conjugated polymer chain networks under elastic stretching evidenced that substantial axial extension is achieved within a few mm from the spinneret,[45,50] and that this depends on the volume fraction occupied by the polymer, $\phi$, the applied electric field, $E$, the jet initial velocity, $v_0$ and the solution conductivity, $K$.[45,51] Typically, lower values of $\phi$ and $v_0$, and higher $E$ and $K$ are needed to increase the axial chain orientation. In our system, the presence of a non-solvent (see Experimental Section) decreases the overall polymer volume fraction in the solution.[41] Moreover, increasing the non-solvent content is expected to improve the solution conductivity as well, because of the two order of magnitude higher electrical conductivity of DMSO compared to chloroform.[52]

This approach is supported by comparing the molecular order of fibers produced from solutions with even slightly different non-solvent content. Figure 2a,b shows the polarized Fourier Transform Infrared (FTIR) spectra of MEH-PPV fibers obtained with chloroform:DMSO mixtures with volume ratios of 9:1 and 9:2, respectively. The spectra show peaks typical of MEH-PPV, such as the modes at 1040 cm$^{-1}$ (ether C-O-C stretch), 1204 cm$^{-1}$ (phenyl-oxygen stretch), 1415 cm$^{-1}$ and 1500 cm$^{-1}$ (C-C ring stretch).[53-55] In both samples, the absorption is maximized for incident light polarization parallel with the fiber longitudinal axis, which is indicative of alignment of the conjugated chains. In particular, the fibers obtained from solutions with higher content of DMSO show a more pronounced difference of spectra measured with the incident light polarized parallel or perpendicular to their axis, meaning that a larger fraction of chain segments are oriented along the fiber length. This is highlighted in Figure 2c, where the amplitude of the peak at 1500 cm$^{-1}$ is shown as a function of the angle, $\theta$, formed by the incident polarization and the direction of fiber alignment (Figure S1a). This mode has a transition dipole moment forming an angle of about 9° with respect to the chain axis,[53] and can be therefore exploited to investigate the degree of chain alignment. An enhanced chain alignment is highlighted by the FTIR analysis upon increasing the DMSO volume fraction up to 20% (Fig. S3a-e), whereas a further increase of the DMSO content negatively affects both the polymer chain alignment and the stability of electrospinning. A decrease of chain alignment is found with DMSO content in the range 20-30%, while continuous solid-state fibers cannot be obtained for DMSO volume fractions above 30%. Overall, these and other[45] results show that electrospinning may be exploited to finely tailor the orientation of conjugated chains in nanostructured materials. Hereafter, fibers obtained by (9:1) and (9:2) content of DMSO will be referred to as "lower molecular order" (LMO) and "higher molecular order" (HMO) samples, respectively.

**X-ray structural analysis**. To probe the molecular orientation and the structure of the HMO samples, free-standing bundles of uniaxially-aligned fibers were investigated by scanning small angle X-ray scattering (SAXS) and wide angle X-ray scattering (WAXS) measurements. Figure 3a shows the scanning SAXS map obtained from a 4×4 mm$^2$ sample area. The contrast variation in the map is related to the amount of material. The predominant color in the 2D map represents the orientation of the anisotropic SAXS signal, and hence of the fibers, according to the color wheel reported in the inset of Fig. 3a.[56] It follows that the greenish color is related to a basically horizontal orientation of the fibers (i.e. parallel to the *x* axis in the map).

The WAXS signal, collected simultaneously and averaged all over the investigated area, is shown in Fig. 3b. Being this rotated by 90° relative to the orientation of Fig. 3a for graphic reasons, the $q_z = 0$ direction (i.e. the $q_y$ axis) in Fig. 3b corresponds to the *y* axis in Fig. 3a. The intensity modulation along the azimuthal angle for the two most intense diffraction rings is plotted in Fig. 3c, where the red and the blue (top) curves are related to the *a* and *c* parameters, respectively (outer and inner rings in Fig. 3b).[57] The azimuthal profiles were extracted from Fig. 3b, at q ~ 1.5 and 1 Å$^{-1}$, respectively, by integrating over a 0.2 Å$^{-1}$ q-range. It can be recognized in Fig. 3c that the intensity modulations of the *a* and *c* parameters are out of phase by 90°, indicating that they are partially aligned along orthogonal preferred orientations. By taking into account the direction of orientations of the fibers as indicated by Fig. 3a, it can be concluded that the *a* and *c* parameters are preferentially oriented perpendicular and parallel to the fiber axis, respectively. Moreover, Fig. 3d shows the azimuthal intensity modulation in the 0.3-0.5 q-range, corresponding to the periodicity of the *b* lattice parameter,[57] which has been here reported in a separated plot because this signal was extracted from WAXS patterns (see Fig. S4a) collected at a larger distance from the sample (120 or 250 mm as specified in Fig. 3d) in order to access lower q-values, although with much lower intensity. Notwithstanding the low signal-to-noise ratio, an overall intensity modulation along the azimuth can be recognized also for the *b* parameter, and well reproduced in both measurements as highligthed by the superimposed smoothed profiles, which is coherent with the azimuthal modulation of the *a* parameter in Fig. 3c. The *b* and *a* parameters are therefore expected to preferentially lie in the plane perpendicular to *c*, and hence to the fiber axis. The WAXS pattern collected at 120 mm sample-to-detector distance is shown in Fig. S4a. In this case, a very thin layer of Si powder (from NIST) was deposited on the bundle of fibers, as an internal standard for calibration and correction of detector tilt.

The preferred orientation is also reflected in the 1D radial profiles of Fig. 3e, resulting from the cuts taken along the horizontal and vertical directions (i.e. perpendicular and parallel to fibers, respectively) in the 2D WAXS pattern of Fig. 3b: the intensity ratio between the two diffraction peaks related to *a* and *c* is seen indeed to be reversed in the two cuts (superimposed curves with symbols and continuous line). A significant fraction of randomly oriented crystalline domains is also expected throughout the fibers, based on the fairly small amplitude of the azimuthal intensity modulation. Moreover, based on the calibration from Fig. S4a, and a Gaussian fit, it was verified that no appreciable shift of peak positions and hence no variations of the related d-spacings occurred between in plane and out of plane directions.

As a result, Fig. 3 indicates the presence of nanocrystals in MEH-PPV electrospun fibers, with crystallographic axes preferentially oriented along the fiber ($c$ axis) or perpendicular to the fiber ($a$ and $b$ axes), determining a preferred orientation of the MEH-PPV active subunits along the fiber axis.

For comparison, MEH-PPV films were studied by collecting the WAXS signal in grazing incidence reflection geometry (see e.g. the GIWAXS pattern in Fig. S4b). By extracting linear cuts from Fig. S4b along the in-plane and out-of-plane directions (1D profiles with black spots and green triangles respectively, in Fig. 3e), an intensity modulation results also in this case, as a difference in the ratios of diffraction peaks related to $a$ and $c$. Such a difference is more evident for the $a$ parameter, which clearly features different diffraction intensities along the two directions. Note that the in-plane profile has been multiplied by a factor 1.1, in order to compensate for the larger absorption because of the larger beam path in the film. Accordingly, in the azimuthal profiles extracted from Fig. S4b and reported in Fig. 3c (bottom curves) for comparison, a corrected intensity larger by a factor 1.1 should be considered, towards the angular interval extremes, in order to compare it with the out-of-plane intensity (corresponding to the center of the curves). Such a correction factor does not significantly affect the azimuthal trends for the $a$ and $b$ parameters (red triangles and black line, respectively, in Fig. 3c), both featuring a clear intensity maximum in the out-of-plane direction. On the contrary, the basically isotropic intensity distribution observed for the $c$ parameter (blue circles) would change to a slightly concave profile, suggesting a slight preferred orientation of the $c$ axis parallel to the sample plane, and hence perpendicular to $a$.

On the other hand, the *b* parameter also features a larger azimuthal intensity perpendicular to the sample plane. Therefore the presence of several differently oriented crystalline domains is expected, particularly with preferred orientation of either the *a* or the *b* axis perpendicular to the sample plane. Overall, a lower degree of crystalline order is found in MEH-PPV thin films compared to electrospun fibers. Indeed, differently from fibers, crystalline domains in films develop independently, with random orientations as well as with the *a* or *b* axis preferentially oriented perpendicular to the film surface, as in a lamellar stacking, whereas the *c* axis of the domains lies preferentially in the sample plane, with random in-plane orientations.

**Steady state optical properties**. MEH-PPV samples with varied degree of chain alignment are investigated by steady state absorption and photoluminescence (PL) spectroscopies. In Figure 4a we compare the absorption spectra of spin cast films and of HMO samples. As a general trend we observe a blue-shift of the absorption peak and a red-shift of the PL peak upon increasing chain alignment (Table I). Interestingly, the absorption peak of the HMO fibers is comparable to the one measured on dilute solutions of MEH-PPV in chloroform (494 nm), suggesting that in fibers the polymer chains adopt a conformation closer to that of a good solvent solution. Furthermore, polarized absorption spectra acquired on arrays of uniaxially aligned HMO fibers show a relative red-shift of the absorption band for polarization of the incident light parallel to the fiber axis (inset of Fig. 4a), which is indicative of a longer conjugation length associated with oriented chromophoric sub-units. The observed spectral shifts are also compatible with an increasing degree of H-aggregation in more ordered samples, according to recent H- and J-aggregate models developed for conjugated polymers.[58,59] In further support of this model, in the film and HMO fiber fluorescence spectra the intensity ratio between the (0-0) and (0-1) vibronic replica, $R_{PL}$=$I_{0-0}$/$I_{0-1}$, is *less* in the HMO fiber (Fig. S5). This ratio provides a quantitative

measurements of the H- or J-aggregation prevalence in conjugated polymers, being typically >2

for dominant intrachain coupling (J-aggregation), whereas values < 2 are indicative of interchain

coupling (H-aggregation).[58,59] Here we find a decrease of $R_{PL}$ from values in the range 1.4-1.7

for films to values of 1.3-1.5 and 1-1.3 for LMO and HMO fibers, respectively, consistent with increased *inter*chain coupling and H-aggregate behavior in fibers. Overall, from polarized FTIR,

X-ray scattering and absorption spectroscopies we argue that during spinning a fraction of the active subchains are effectively stretched and oriented along the fiber axis, adopting a more extended chain conformation compared to amorphous films, with an increase of the associated conjugation length and a packing arrangement that supports interchain interactions with nearby subchains (Figure 4b,c).

More importantly, the fibers with more oriented polymer chains feature an increase of the PL

quantum yield ($\Phi_F$) by about a factor 5 compared to films (Table I). These results are in striking

contrast to what is typically observed in conjugated polymers in the solid state, where a red-shift

of the emission is associated with the formation of low-emissive aggregates.[60]

To investigate more in depth the origin of the enhanced and red-shifted emission, we perform a

study of the polarization properties of a set of individual nanofibers by micro-photoluminescence

(μ-PL). Figures 5 and 6 display the polarized PL spectra of films and individual LMO and HMO

fibers, acquired by exciting the fibers with a laser polarized either parallel with (Fig. 5) or perpendicular to (Fig. 6) their length. Independent of excitation configuration, the emission from films is largely unpolarized, as is characteristic of systems with randomly oriented chains, while the emission from fibers is polarized along their length (Fig. 5a-c and 6a-c). The degree of molecular alignment affects the amount of polarized light detected, since the ratio between the intensity of the PL polarized parallel with ($I_\parallel$) and perpendicular ($I_\perp$) to the fiber axis ($\chi = I_\parallel / I_\perp$) increases by enhancing the chain orientation (Fig.s 5, 6, S3f and Table I). Moreover, the emission polarized parallel to the fiber longitudinal axis is red-shifted up to about 6 nm compared to the perpendicular one, and the spectral shift is even more pronounced for excitation polarization perpendicular to the fiber axis (Fig. S6 and Table I). While in films energy migration is effective in randomizing the emission polarization, which causes any memory of the excitation polarization to be lost, in nanofibers energy migration funnels energy towards chromophores oriented along the fiber axis, which emit at lower energy (i.e., are red-shifted) as a consequence of their more extended and conjugated network. The high stretching rate and electric field applied during the spinning process causes a prevalent orientation of chains along the jet axis but also their disentanglement,[61,62] favoring H-aggregation and interchain interactions.

The increased emission quantum yield is therefore attributable to the unique processing conditions that, in contrast to other solution-based deposition processes, freeze the polymer chains in highly anisotropic configurations yet retain strong interchain interactions that promote energy migration. Remarkably, the emission polarization found on HMO individual fibers is indeed comparable to the PL polarization measured in isolated MEH-PPV chains,[20,26] i.e. the ultimate achievable polarization properties for the emission of the studied system.

**Table I**. Steady state optical properties of MEH-PPV films and fibers with low (LMO) and high (HMO) molecular order. $\chi_{//}$ ($\chi_{\perp}$) is the ratio between the PL intensity polarized parallel and perpendicular to the fiber axis, collected upon excitation with a laser polarized parallel with (perpendicular to) the fiber axis. The associated errors are obtained as the standard deviation of the distributions shown in Figure 5 and 6. $\Delta_{//}$ ($\Delta_{\perp}$) is the spectral shift between the PL polarized parallel and perpendicular to the fiber axis, upon excitation with a laser polarized parallel with (perpendicular to) the fiber axis.

| | Film | LMO Nanofibers | HMO Nanofibers |
|---|---|---|---|
| Abs $\lambda_{max}$ (nm) | 500±1 | 497±1 | 492±1 |
| PL $\lambda_{max}$ (nm) | 578±1 | 590±1 | 598±1 |
| $\Phi_F$ | 0.05±0.01 | 0.18±0.01 | 0.23±0.01 |
| $\chi_{//}$ | 1.0±0.1 | 2.0±0.3 | 4.4±0.7 |
| $\Delta_{//}$ (nm) | 0.6±0.1 | 2.6±0.1 | 2.2±0.2 |
| $\chi_{\perp}$ | 1.0±0.1 | 1.7±0.3 | 3.8±0.5 |
| $\Delta_{\perp}$ (nm) | 0.4±0.1 | 5.0±0.1 | 5.8±0.1 |

**Pump–probe stimulated emission anisotropy.** To investigate the origin of the anisotropic emission properties of the MEH-PPV films and fibers and more clearly understand energy migration in these systems, we perform pump-probe stimulated emission anisotropy measurements (Fig. 7). The stimulated emission feature in pump-probe anisotropy measurements has been shown to be a sensitive probe of the migration of energy in conjugated polymers.[20] In this technique, a pump beam with a defined linear polarization state prepares a sub-ensemble population of excitons whose transition dipole moments are aligned parallel to the laser beam. A probe beam then interrogates the state of the system with a polarization state that is variably oriented either parallel or perpendicular to that of the pump beam. The measured pump-probe stimulated emission data can be used to calculate the time-dependent anisotropy of excitons in the system, that is, their polarization memory, according to the following equation:

$$r(t) = (S_{para}(t) - S_{perp}(t))/(S_{para}(t) + 2S_{perp}(t)) \qquad (1)$$

The anisotropy of the 0-0 band of the stimulated emission of the MEH-PPV film decays on the picosecond timescale toward a value close to zero. In the experiments on the MEH-PPV nanofibers, the nanofiber axis was oriented perpendicular to the pump beam polarization. We find that the anisotropy of the 0-0 band of the stimulated emission of the MEH-PPV HMO nanofibers also decays on the picosecond timescale, but in contrast to the MEH-PPV film, continues to decay beyond zero taking on negative values at long pump-probe time delays.

We observe that the polarization memory of excitons in both films and HMO nanofibers of MEH-PPV is lost on the few picosecond timescale (Fig. 8). A stretched exponential function with a time constant of 2 ps reasonably well describes the anisotropy decay of the MEH-PPV films, and overlay the data in Figure 8 as a guide to the eye. These observations are generally consistent with prior measurements on MEH-PPV films.[42] In the MEH-PPV HMO nanofibers, a stretched exponential function with a time constant of 1 ps adequately describes the anisotropy decay. Given the relatively fast depolarization timescale, this suggests a prominent role of interchain electronic energy transfer in both systems.[21,63] Significantly, whereas the decay of the anisotropy of the 0–0 band of the stimulated emission in the MEH-PPV film approaches zero at long pump–probe time delay, the corresponding decay in the MEH-PPV nanofibers takes on negative values at long pump–probe time delay. These observations suggest a preferential alignment of excitons along the fiber axis following energy migration.

### 3.  CONCLUSIONS

In summary, MEH-PPV fibers with oriented polymer chains are investigated by modulating the composition of the solution used for electrospinning. The structural characterization by X-ray diffraction measurements highlights a preferred orientation of molecules in the electrospun fibers, with the $c$ axis parallel to the fiber axis and the $a$ and $b$ axes perpendicularly oriented, evidencing the formation of a well-defined  3-dimensional nanocrystalline order, differently from films, where randomly oriented crystalline domains are present as well as domains with different preferred orientations. The anisotropic packing and stretching of the MEH-PPV macromolecules in the solid state fibers leads to a five-fold enhancement of the PL quantum yield compared to samples composed of chains exhibiting disordered packing (i.e., spin cast films). Polarized steady state spectroscopy and femtosecond pump–probe anisotropy measurements indicate that energy migration prevalently funnels excitations towards chromophores oriented along the fiber axis, within a characteristic picosecond timescale. These emissive chromophores feature longer conjugation lengths and increased interchain interactions, demonstrating improved emission properties of the fibers compared to films. These nanostructures with precisely tailored internal molecular order are an example of how more efficient and directed energy transfer can be achieved, and can potentially enhance performance in light-harvesting architectures.

## 4. EXPERIMENTAL SECTION

*Nanofibers*. MEH-PPV (M.W. 380,000 g/mol) was purchased from American Dye Source, Inc. and used as received. For electrospinning experiments, 36 mg of MEH-PPV were dissolved in 1 mL of a $CHCl_3$:DMSO mixture, with relative ratios in the range 9:1 to 9:4 (v:v). Experiments performed with higher content of DMSO did not allow a stable electrospinning jet and uniform solid state fibers to be obtained. The solution was stirred for 24 h to achieve the complete dissolution of the conjugated polymer. Afterwards, the solution was stored in a 1 mL syringe, tipped with a 27 G stainless steel needle. The syringe was loaded in a microprocessor dual drive syringe pump (33 Dual Syringe Pump, Harvard Apparatus), used with a constant rate of 0.5 mL/h. A rotating disk (4000 rpm) positioned 10 cm far away from the needle was used as a collector, allowing uniaxially aligned arrays of fibers to be obtained. All experiments were performed by applying a positive bias of 10 kV to the needle, while the collector was biased at a negative voltage of -5 kV. Quartz coverslip (thickness 250 μm) were used as substrates. Following deposition, fibers were embedded in a photocurable polymer (NOA68, Norland Products Inc., refractive index, *n*=1.54), which was cured for 3 min. The thickness of the cured resin was about 1 mm. Alternatively, free-standing bundles of uniaxially-aligned fibers are produced by using the rotating collector and punched Al foils as substrates. Reference films were produced by spin coating at 6000 rpm on quartz substrates for 40 seconds. The reference dilute solution was produced by dissolving 0.2 mg of MEH-PPV powder in 1 mL of $CHCl_3$.

The morphology of the fibers was investigated by scanning electron microscopy (SEM) and atomic force microscopy (AFM). SEM was performed by a Nova NanoSEM 450 system (FEI), operating at an acceleration voltage of 3 kV. AFM was performed by a Multimode head, equipped with a Nanoscope IIIa electronic controller (Veeco Instruments), imaging in tapping mode with Silicon cantilevers (resonance frequency 250 kHz). Polarized FTIR spectroscopy was carried out by using a spectrophotometer (Spectrum 100, Perkin Elmer Inc.), equipped with a IR grid polarizer (Specac Limited, UK), consisting of 0.12 μm-wide strips of aluminum, here used to generate an incident polarized light beam. Fluorescence imaging of the fibers arrays was carried out by using an inverted microscope (Eclipse Ti, Nikon) and an A1R MP confocal system (Nikon). The samples were excited by an $Ar^+$ ion laser ($\lambda_{exc}$ = 488 nm) through an oil immersion objective with numerical aperture (N.A.) of 1.4, which also collected the light emitted by the sample. The fluorescence intensity and spectral distribution was measured by a spectral detection unit equipped with a multi-anode photomultiplier (Nikon).

*X-ray diffraction measurements*. SAXS, WAXS and GIWAXS measurements were performed on a laboratory (GI)SAXS/(GI)WAXS set-up (XMI-Lab),[64] equipped with a Rigaku Fr-E+ Superbright microsource ($\lambda$= 0.154 nm) and a SMAX3000 three pinhole camera. A 200 μm (diameter) X-ray beam was employed. SAXS patterns were collected at a ~2.2 m distance from the sample, with a 200 μm step size, by using a multiwire Triton detector. GIWAXS patterns were collected at a 0.2° incidence angle, on an image plate (IP) detector with 100 μm pixel size at a 87 mm sample-to-detector distance. Patterns - SAXS/(GI)WAXS - collected at 87 mm, or more, from the sample were calibrated by using a standard Silver Behenate powder sample, or a Si powder from NIST as specified in the text. WAXS patterns collected at a 28 mm distance were calibrated by using a standard LaB6 powder from NIST. The batch for the scanning

SAXS/WAXS experiment and the composition of data in color map were realized by using the in-house software SUNBIM.[65]

*Optical spectroscopy*. Absorption spectra were collected by using a Tungsten lamp and a monochromator equipped with a CCD camera detector (USB4000, Ocean Optics). The samples were mounted in an integrating sphere (Labsphere), in order to minimize artifacts due to light scattering. Polarized absorption spectra were obtained by using incident light polarized either parallel or perpendicular to the fiber axis. Similarly, PL spectra were acquired exciting the samples, placed in the integrating sphere, by a laser (Coherent VERDI, $\lambda_{exc}$=532 nm, beam diameter = 1 mm) and collecting the emission by an optical fiber coupled to a monochromator. This system was also used for measuring the PL quantum yield following the procedure reported in Ref. 66.

*Polarized micro-photoluminescence*. The polarization of individual fibers was investigated by a μ-PL system. This system includes a linearly polarized diode laser ($\lambda_{exc}$=405 nm), focused onto single fibers by a 20× objective (N.A.=0.5) of an inverted microscope (IX71, Olympus). The vertically emitted light was then collected by a spherical lens ($f$=60 mm), coupled to an optical fiber and measured by a spectrometer. The polarization state of the emission was analyzed by a polarization filter, which was placed on a precision rotation stage, between the sample and the collecting lens. The samples were placed on a rotation stage, which allows the fiber axis to be precisely aligned with the polarization direction of the excitation laser.

*Pump-probe stimulated emission anisotropy.* The pump-probe stimulated emission anisotropy measurements were performed on a pump-probe spectrometer that has been described in detail previously.[67] Briefly, a Ti:sapphire-based regenerative amplifier (Spectra-Physics Spitfire Pro) generates about 3.5 W of 800 pulsed laser light that has a duration of ~100 fs and a repetition rate of 5 kHz. A portion of the output is directed toward a custom-built noncollinear optical parametric amplifier (NOPA) that converts the 800 nm radiation into wavelengths in the visible region. In the present experiments, the NOPA was tuned to encompass a spectral range from about 500 to 620 nm. The output of the NOPA is directed towards a series of compressors, including a grating and prism compressor. Details of the optical layout of the NOPA as well as the grating and prism compressors were reported previously.[68] The NOPA spectrum is shown in Figure S7. The output of the prism compressor is then directed towards the pump-probe spectrometer where a glass wedge serves to split the beam into pump and probe beam paths. Both pump and probe beams pass through a waveplate and Glan-Thompson polarizer (New Focus), the combination of which allows to variably control the beam intensity and define the polarization state of the beams at the sample position. The path length of the pump beam is variably controlled with an automated, mechanical delay stage (Newport, Santa Clara, CA). The probe beam transmitted through the sample is collected and directed toward a monochromator and camera combination (Andor) for spectral detection. An optical chopper (New Focus) in the pump beam path operating at 625 Hz enables differential detection. Balanced detection is achieved by simultaneously accounting for fluctuations in the laser beam by using a photodiode to monitor the intensity of the second beam reflected from the glass wedge. Pulse compression was guided by minimizing the full-width-at-half-maximum of the coherent artifact measured from a solution of methanol in a spectrophotometer cell placed at the sample position that was subsequently replaced with a blank resin substrate. Pump pulse energies were determined by measuring the beam power and the beam spot sizes were estimated by comparing the power transmitted through a pinhole placed at the sample position. Incident pump energy densities estimated in this manner were ~40 and ~15 $\mu J/cm^2$ in the film and nanofiber measurements, respectively.

## ASSOCIATED CONTENT

**Supporting Information**. Additional data on nanofiber alignment analysis, photostability, polarized FTIR analysis, X-ray scattering measurements, photoluminescence spectra, polarized emission spectra and laser pulse spectrum for the pump-probe anisotropy measurements. This material is available free of charge via the Internet at http://pubs.acs.org.

## AUTHOR INFORMATION

**Corresponding Authors**

*Andrea Camposeo. E-mail address: andrea.camposeo@nano.cnr.it

**Dario Pisignano. E-mail address: dario.pisignano@unisalento.it

*** Gregory D. Scholes. E-mail address: gscholes@princeton.edu

## ACKNOWLEDGMENT

Rocco Lassandro is acknowledged for his technical support in the XMI-Lab. The research leading to these results has received funding from the European Research Council under the European Union's Seventh Framework Programme (FP/2007-2013)/ERC Grant Agreement n. 306357 (ERC Starting Grant "NANO-JETS"). The Apulia Networks of Public Research Laboratories WAFITECH (09) and M. I. T. T. (13) are also acknowledged.

FIGURES AND CAPTIONS

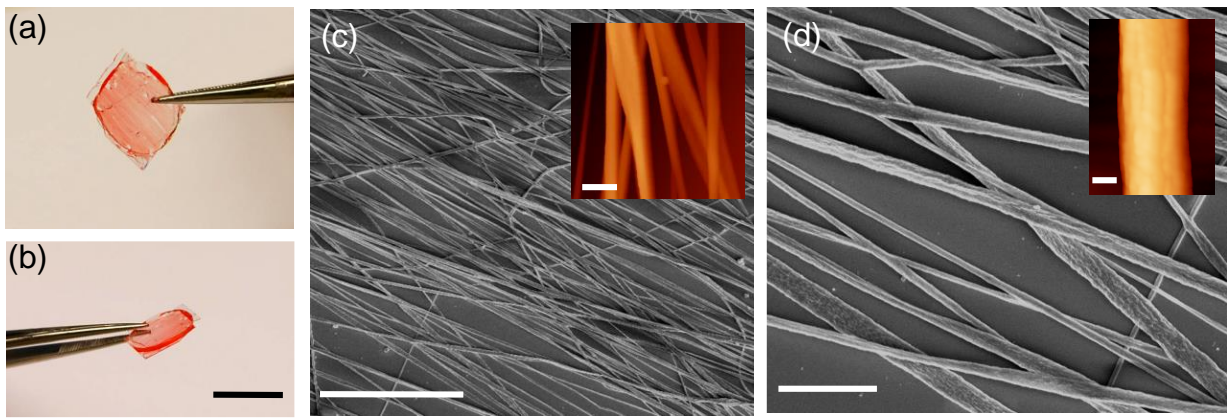

**Figure 1**. **MEH-PPV nanofiber morphology**. (a)-(b) Photographs of MEH-PPV fibers encapsulated in a photocurable resin. Scale bar: 1 cm. (c)-(d) SEM images of the electrospun fibers. Scale bars: (c) 30 μm and (d) 5 μm. Insets: corresponding AFM images, highlighting the fiber surface morphology. Inset scale bars: (c) 2 μm and (d) 100 nm.

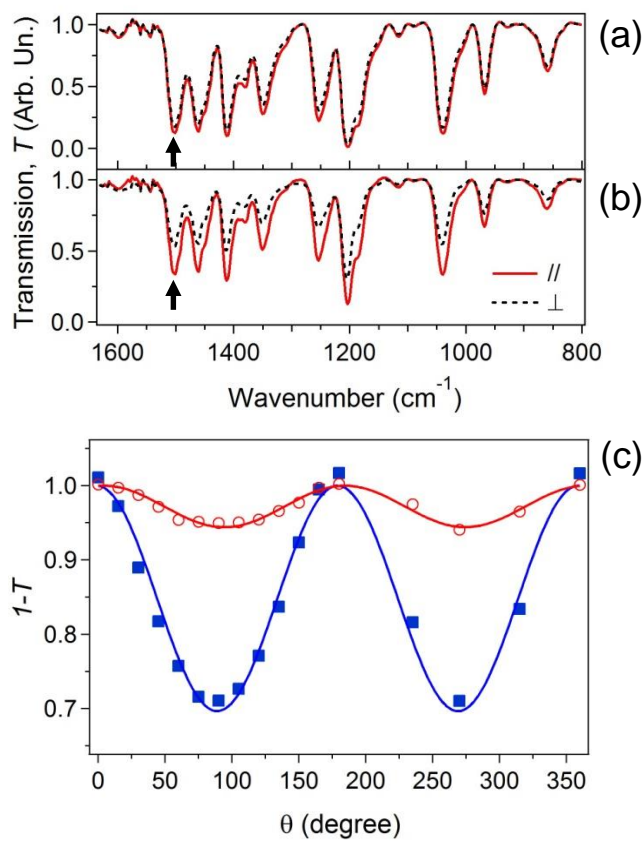

**Figure 2**. **Analysis of polymer chain alignment by FTIR polarized spectroscopy**. (a)-(b) Polarized FTIR spectra of uniaxially aligned MEH-PPV fibers spun from chloroform:DMSO mixtures with volume ratio of 9:1 (a) and 9:2 (b). The spectra are obtained by using incident light with polarization parallel (red continuous line) or perpendicular (black dashed line) to the fiber alignment direction. The arrows indicate the mode at about 1500 cm$^{-1}$, attributed to a C-C ring stretching mode. (c) Amplitude of the transmission peak at 1500 cm$^{-1}$ *vs*. the angle, θ, formed by the fiber axis and the incident polarization direction (see Figure S1 of the Supporting Information) for fibers spun from chloroform:DMSO mixtures with volume ratio of 9:1 (circles) and 9:2 (squares). Data are normalized to their maximum values. 0° corresponds to the direction of the incident light polarization parallel to the fiber axis. The continuous lines are fits to the data by a cos$^2$ law.

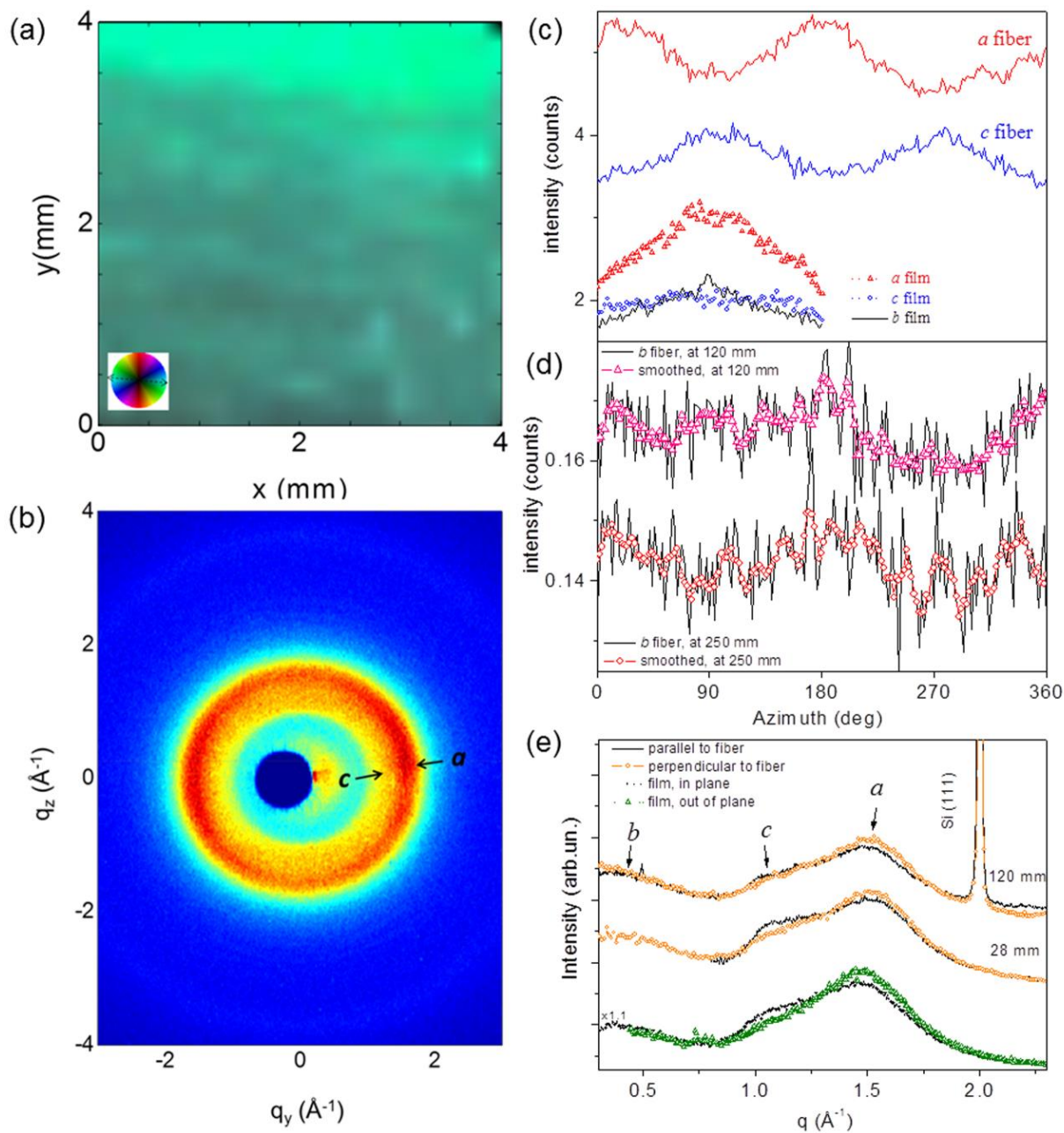

**Figure 3**. **X-ray diffraction analysis**. Simultaneous (a) scanning SAXS microscopy and (b) integrated WAXS microscopy of a bundle of uniaxially-aligned free-standing fibers. (c) Top: azimuthal profiles along the rings of Fig. 3b at q~1 and q~1.5 Å⁻¹, related to the $c$ and $a$ lattice parameters, respectively. Bottom: azimuthal profiles along the rings of Fig. S4b at q~0.4, q~1 and q~1.5 Å⁻¹, related to the $b$, $c$ and $a$ lattice parameters, respectively (data multiplied by a factor 2 for better clarity). (d) Azimuthal profiles taken at q~0.4 Å⁻¹ for the $b$ lattice parameter of the free-

standing fibers, from 2D SAXS patterns collected at 120 mm (Fig. S4a) and 250 mm (not shown).

(e) Linear cuts taken from the WAXS patterns of Fig. 3b (28 mm detector distance) and S4a (120 mm detector distance), in the directions parallel (black line) and perpendicular (orange symbols) to the fiber axis, and from the GIWAXS pattern of Fig. S4b in the directions parallel (black dots) and perpendicular (green triangles) to the film surface. The in-plane profile has been multiplied by a factor 1.1, in order to compensate for the larger absorption due to the larger beam path in the film.

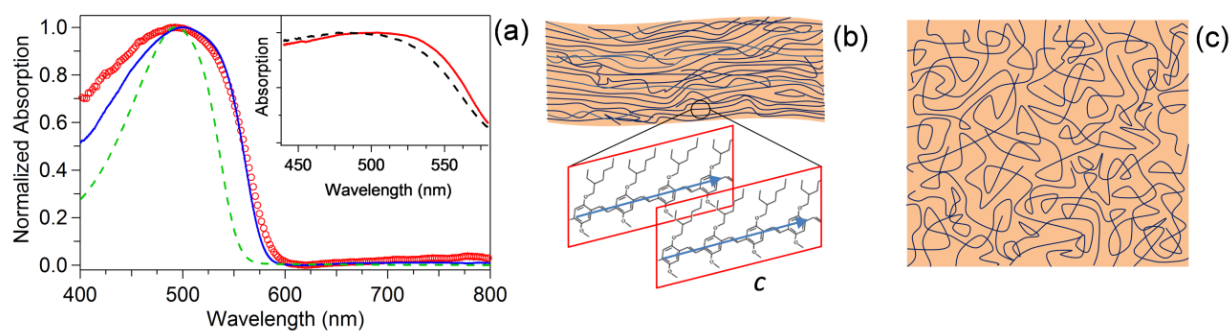

**Figure 4**. **Steady state optical absorption**. (a) Absorption spectra of a MEH-PPV spin-cast film (continuous line) and HMO fibers (circles). The films are produced by using the solution of HMO fibers. The spectra are normalized to their maxima. The absorption spectrum of a dilute solution of MEH-PPV in chloroform is shown for reference (dashed line). Inset: polarized absorption spectra of HMO fibers, acquired by incident light polarized parallel (continuous line) or perpendicular (dashed line) to the fiber axis. (b)-(c) Pictorial representations of the microscopic arrangement of individual conjugated polymer chains in nanofiber (b) and film (c). The inset in (b) highlights the polymer sub-chain conformation. Here, arrows indicate the *c* axis of the crystalline domains, whereas *a* and *b* axes lie in a plane perpendicular to *c*.

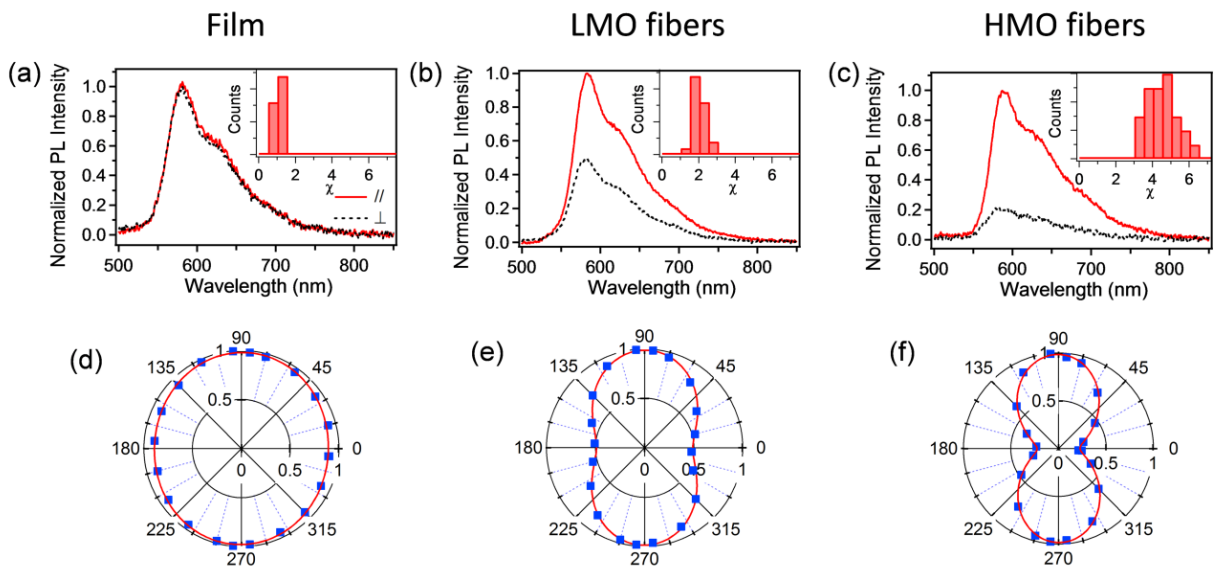

**Figure 5**. **Polarization PL analysis with excitation polarization parallel to fiber length**. (a) PL spectral components of MEH-PPV films polarized either parallel with (red continuous lines) or perpendicular (black dashed lines) to the excitation laser polarization. (b)-(c) PL spectra of MEH-PPV fibers polarized either parallel with (red continuous lines) or perpendicular (black dashed lines) to the fiber longitudinal axis. Samples are excited with a laser polarized parallel with the fiber axis. The insets show the distributions of the nanofiber polarization ratio, $\chi$, obtained by measuring several individual MEH-PPV nanofibers, or different regions of the film. (d)-(f) Plot of the sample emission intensity vs. the angle between the fiber and the polarization axis of the collection filter. The emission intensity maximum corresponds to the axis of polarization filter parallel with fiber length. Continuous lines are fits to the data by a $\cos^2$ law. (a) and (d): films; (b) and (e) LMO fibers; (c) and (f) HMO fibers.

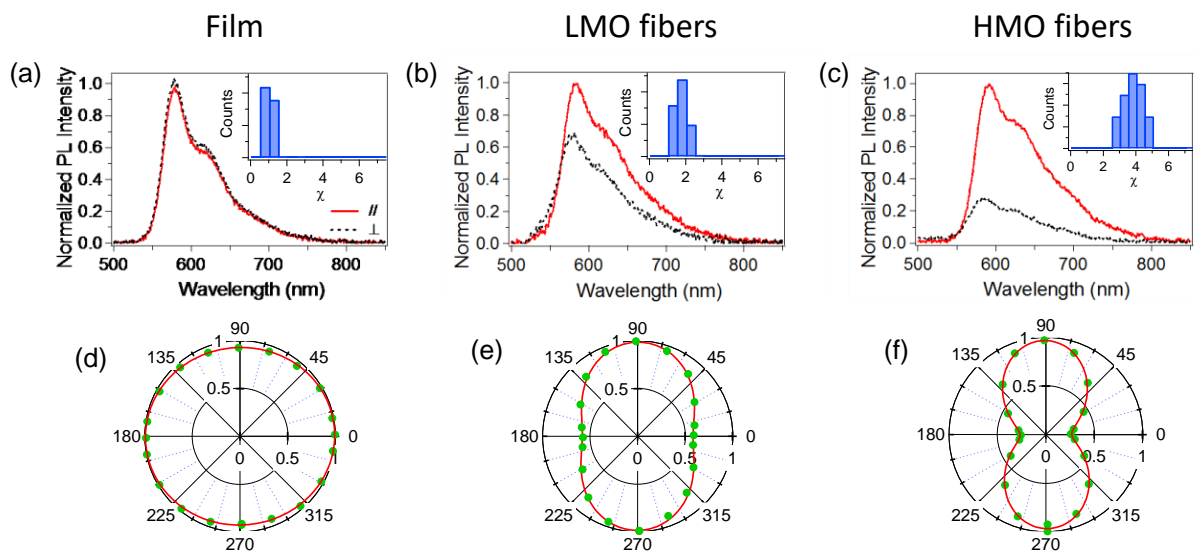

**Figure 6**. **Polarization PL analysis with excitation polarization perpendicular to fiber axis**. (a) PL spectral components of MEH-PPV films polarized along two orthogonal directions. (b)-(c) PL spectra of MEH-PPV fibers polarized either parallel with (red continuous lines) or perpendicular (black dashed lines) to the fiber longitudinal axis. Samples are excited with a laser polarized perpendicular to the fiber axis. The insets show the distributions of the nanofiber polarization ratio, χ. (d)-(f) Plot of the sample emission intensity vs. the angle between the fiber and the polarization axis of the collection filter. The emission intensity maximum corresponds to the axis of polarization filter parallel with fiber length. Continuous lines are fits to the data by a $\cos^2$ law. (a) and (d): films; (b) and (e): LMO fibers; (c) and (f): HMO fibers.

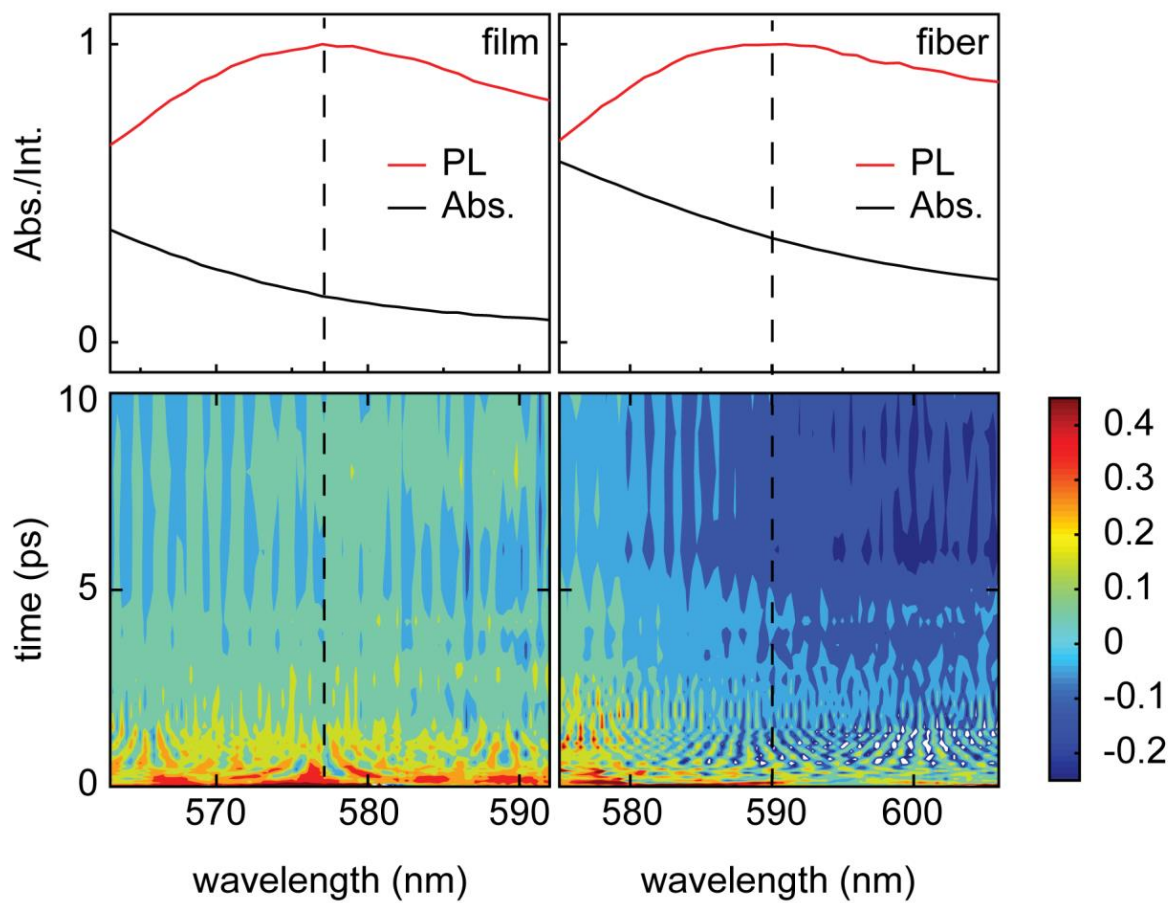

**Figure 7. Pump–probe anisotropy measurements of an MEH-PPV film and HMO nanofibers.** Steady-state absorption and photoluminescence of an MEH-PPV film and HMO nanofibers (upper panels). These PL measurements were made with $\lambda_{exc}$ = 500 nm. Pump–probe anisotropy decay observed in the vicinity of the 0–0 band of the stimulated emission of an MEH-PPV film and HMO fibers (bottom panels). The data are plotted with seven counters in intervals of 0.1 over the range -0.25 to 0.45. The incident pump energy density in the case of the film and nanofiber measurements were ~40 and ~15 μJ/cm², respectively.

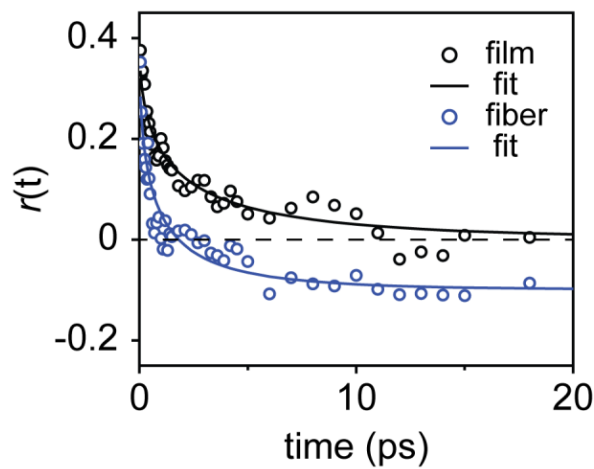

**Figure 8. Pump–probe anisotropy decay of the 0–0 band of the stimulated emission of an MEH-PPV film and HMO nanofibers.** The decay of the 0–0 band of the stimulated emission was taken as the mean over the spectral range 563 to 592 nm and 575 to 606 nm for the MEH-PPV film and HMO nanofibers, respectively. Stretched exponential functions with time constants of ~2 and ~1 ps corresponding to the decay of the anisotropy of the stimulated emission observed in the MEH-PPV film and nanofibers, respectively, overlay the data as a guide to the eye.

# SUPPORTING INFORMATION

# Anisotropic conjugated polymer chain conformation tailors the energy migration in nanofibers

Andrea Camposeo[†*], Ryan D. Pensack[#], Maria Moffa[†], Vito Fasano[§], Davide Altamura[‡], Cinzia Giannini[‡], Dario Pisignano[†,§**], Gregory D. Scholes[#***]

[†] Istituto Nanoscienze-CNR, Euromediterranean Center for Nanomaterial Modelling and Technology (ECMT), via Arnesano, I-73100, Lecce, Italy.

[#] Department of Chemistry, Princeton University, Princeton NJ 08544 U.S.A.

[‡] Istituto di Cristallografia (IC-CNR), via Amendola 122/O, I-70126, Bari, Italy.

[§] Dipartimento di Matematica e Fisica "Ennio De Giorgi", Università del Salento, via Arnesano, I-73100, Lecce, Italy.

*Andrea Camposeo. E-mail address: andrea.camposeo@nano.cnr.it

**Dario Pisignano. E-mail address: dario.pisignano@unisalento.it

*** Gregory D. Scholes. E-mail address: gscholes@princeton.edu

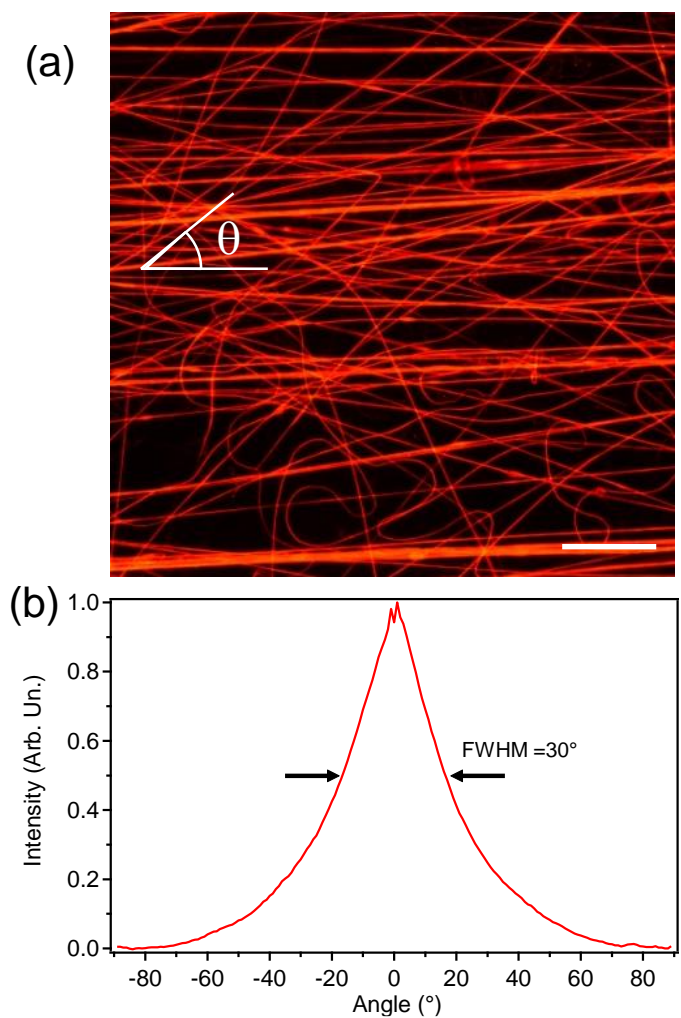

**Figure S1**. **Analysis of fiber alignment**. (a) Confocal fluorescence micrograph of aligned MEH-PPV fibers. Scale bar: (a) 50 μm. (b) Distribution of fiber axis alignment angles, calculated from data shown in (a). The angle reported in the *x*-axis of panel (b) corresponds to the angle, θ, schematized in (a). These data evidence the alignment of electrospun fibers in an angular range of about 30°.

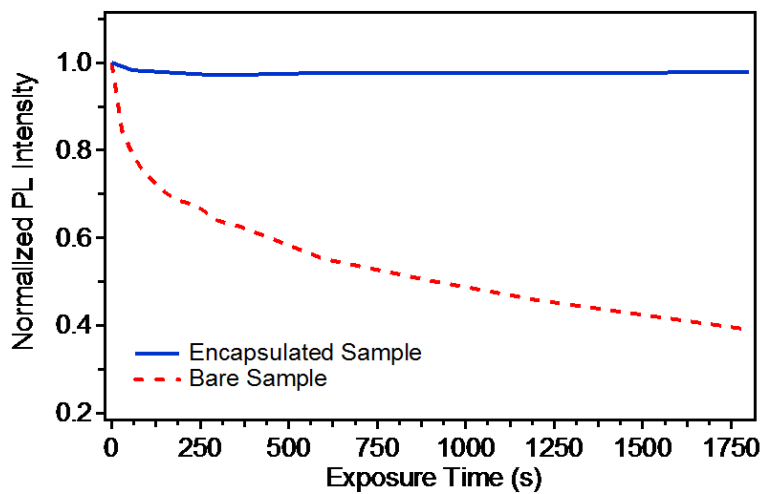

**Figure S2**. **Photostability of encapsulated fibers**. Temporal decay of the photoluminescence (PL) intensity upon continuous exposure to an excitation laser beam ($\lambda_{exc}$=410 nm, incident intensity 20 mW cm$^{-2}$), for a time interval of 30 minutes. Data show an almost unchanged PL intensity for the encapsulated fibers (continuous line), whereas the PL intensity of the MEH-PPV fibers in air (dashed line) is halved in about 15 minutes.

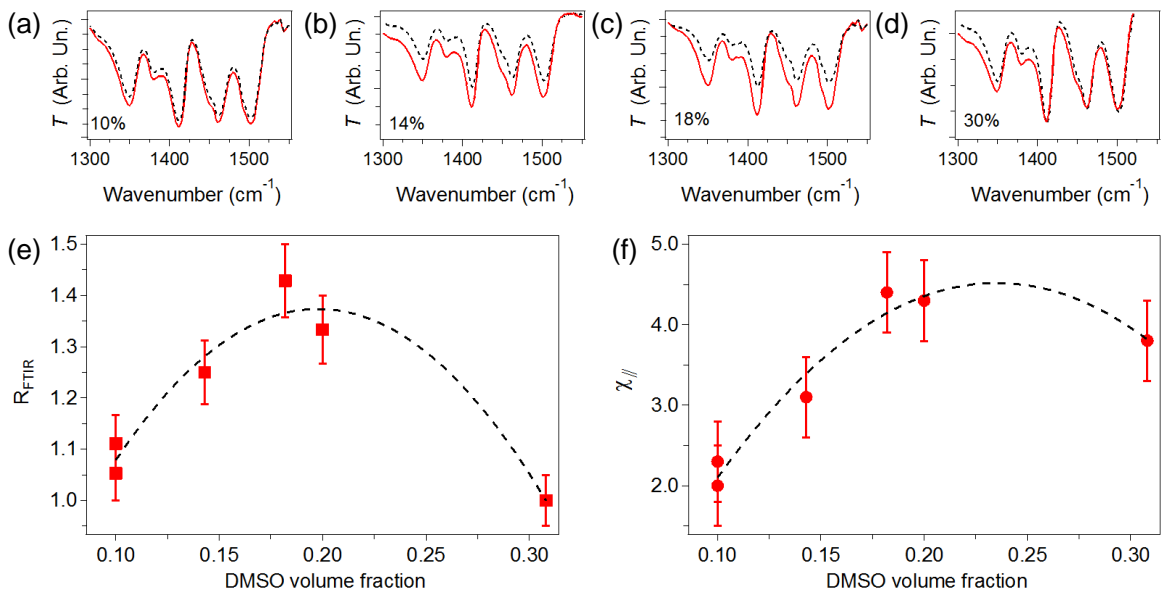

**Figure S3**. **Analysis of the structure anisotropy as a function of processing conditions.** (a)-(d) Polarized FTIR spectra of uniaxially aligned MEH-PPV fibers spun from chloroform:DMSO mixtures with a DMSO volume fraction of 10% (a), 14% (b), 18% (c) and 30% (d). The spectra are measured by using incident light with polarization parallel (red continuous line) or perpendicular (black dashed line) to the fiber alignment direction. (e) Dependence of the degree of chain alignment vs. content of DMSO. The plotted data are calculated as $R_{FTIR}=A_{///} A_{\perp}$, where $A_{//}$ and $A_{\perp}$ are the amplitudes of the peak at 1500 cm$^{-1}$ of the FTIR spectra collected with polarization of the incident beam parallel and perpendicular to the fiber length, respectively. (f) Dependence of $\chi_{//}$ on the content of DMSO. $\chi_{//}$ is the ratio of the intensity of PL, polarized parallel and perpendicular to the fiber axis, collected upon excitation with a laser polarized parallel with the fiber axis.

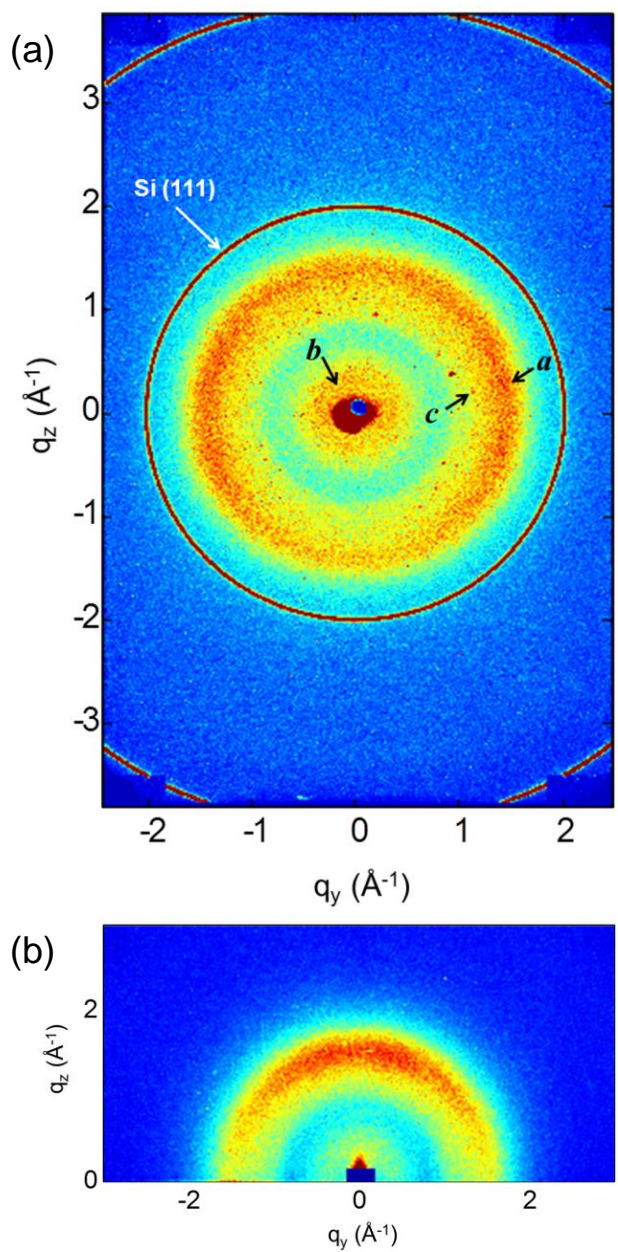

**Figure S4**. **X-Ray diffraction patterns**. (a) Transmission WAXS pattern of the free-standing

fibers sample, with internal Si powder standard, collected at a 120 mm sample-detector distance.

(b) Grazing incidence WAXS pattern of the thin MEH-PPV film.

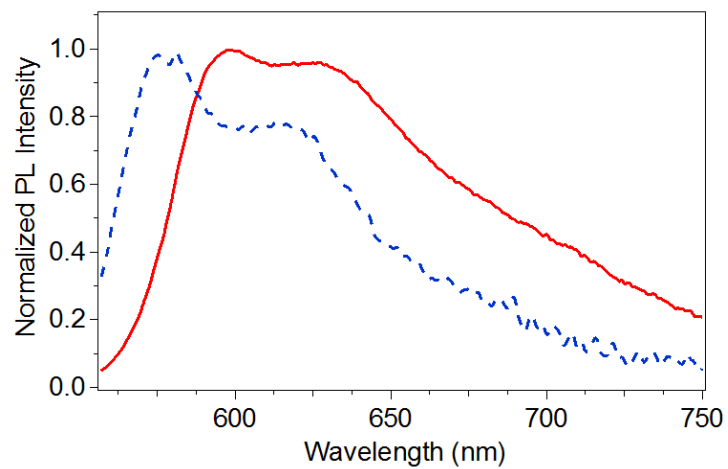

**Figure S5**. **Steady-state photoluminescence spectra**. Photoluminescence spectra of a MEH-PPV spin-cast film (dashed line) and HMO fiber mat (continuous line). The films are produced by using the solution of HMO fibers. The spectra are acquired with and integrating sphere (see Methods) and normalized to their maxima.

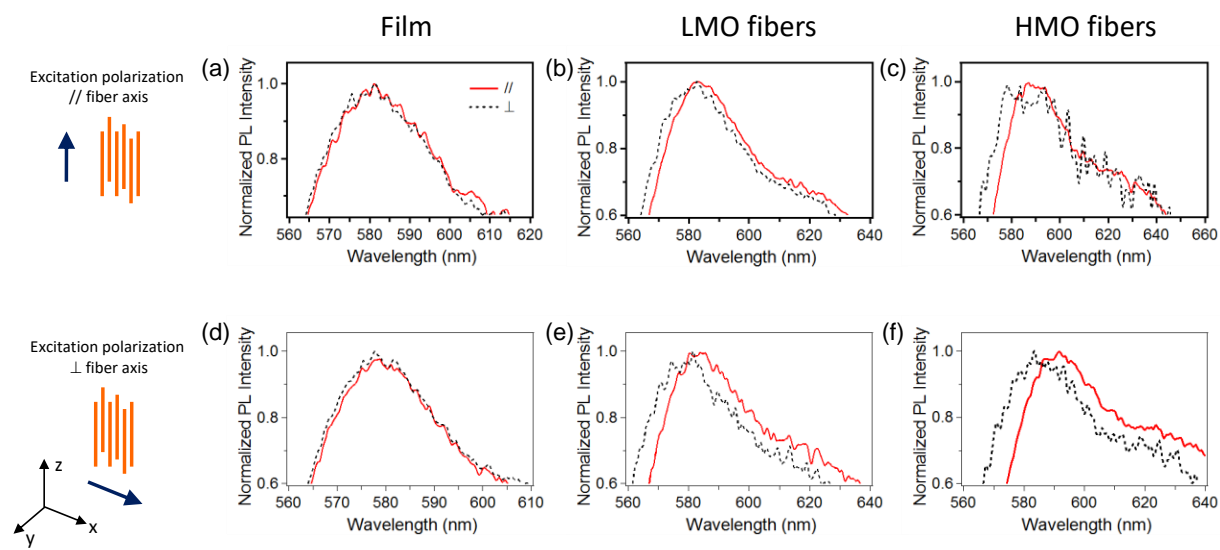

**Figure S6**. Normalized polarized PL spectra obtained by collecting the emitted light through a polarization filter with axis aligned along two orthogonal directions. For fiber samples, the polarization filter had axis parallel (red continuous lines) and perpendicular (black dashed lines) to the fiber axis. The spectra are collected by exciting the samples with the polarization direction of the excitation laser parallel (b)-(c) and perpendicular (e)-(f) to the fiber axis, as schematically depicted on the left. The spectra are normalized to their respective maxima.

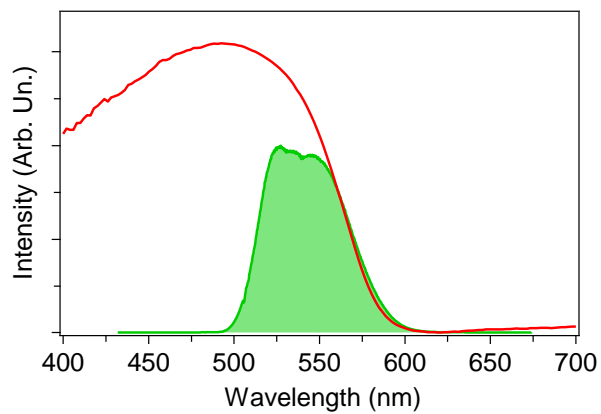

**Figure S7**. Spectral profile of the excitation laser (green line) used for the temporal transient measurements. As a reference, the absorption spectrum of the MEH-PPV HMO fibers is shown (red continuous line).